\documentstyle[12pt,aaspp4]{article}
\tighten
\renewcommand{\baselinestretch}{1.5}
\newbox\grsign \setbox\grsign=\hbox{$>$} 
\newdimen\grdimen \grdimen=\ht\grsign
\newbox\laxbox \newbox\gaxbox
\setbox\gaxbox=\hbox{\raise.5ex\hbox{$>$}\llap
     {\lower.5ex\hbox{$\sim$}}}\ht1=\grdimen\dp1=0pt
\setbox\laxbox=\hbox{\raise.5ex\hbox{$<$}\llap
     {\lower.5ex\hbox{$\sim$}}}\ht2=\grdimen\dp2=0pt

\def\ls{LS~5039}

\def\ltsima{$\; \buildrel < \over \sim \;$}
\def\simlt{\lower.5ex\hbox{\ltsima}}            % < over MMM
\def\gtsima{$\; \buildrel > \over \sim \;$}
\def\simgt{\lower.5ex\hbox{\gtsima}}            % > over MMM

\begin{document}

\title{Science 288, 2340, 30 June 2000}

\vspace{0.5cm}

\title{Discovery of a High-Energy Gamma-Ray-Emitting Persistent 
Microquasar}

\author{Josep M. Paredes,\altaffilmark{1*}
Josep Mart\'{\i},\altaffilmark{2}
Marc Rib\'o,\altaffilmark{1}
Maria Massi\altaffilmark{3} 
}

\vspace{2cm}

\noindent
\altaffilmark{1}{Departament d'Astronomia i Meteorologia, Universitat
de Barcelona, Av. Diagonal 647,
E-08028 Barcelona, Spain} 

\noindent
\altaffilmark{2}{Departamento de F\'{\i}sica, Escuela Polit\'ecnica
Superior, Universidad de Ja\'en, Calle Virgen de la Cabeza 2, E-23071
Ja\'en, Spain} 

\noindent
\altaffilmark{3}{Max Planck Institut f\"ur Radioastronomie, Auf dem
H\"ugel 69, D-53121 Bonn, Germany} 

\vspace{1cm}

\noindent
\altaffilmark{*}{To whom correspondence should be addressed. E-mail:
josepmp@am.ub.es}

\newpage

\noindent
Microquasars are stellar x-ray binaries that behave as a scaled down
version of extragalactic quasars. The star \ls\ is a new microquasar
system with apparent persistent ejection of relativistic plasma at a 3
kiloparsec distance from the sun. It may also be associated with a
$\gamma$-ray source discovered by the Energetic Gamma Ray Experiment
Telescope (EGRET) on board the COMPTON-Gamma Ray Observatory satellite.
Before the discovery of \ls, merely a handful of microquasars had been
identified in the Galaxy, and none of them was detected in high-energy
$\gamma$-rays.

\newpage

\noindent
The $V=11.2$ magnitude star \ls\ ({\it 1}) has been recently identified as
a nearby high-mass x-ray binary with spectral type O7V((f)) ({\it 2}) and
persistent radio emission ({\it 3,4}). Here, we report high-resolution
radio observations with the Very Long Baseline Array (VLBA) and the Very
Large Array (VLA) that reveal that \ls\ is resolved into bipolar radio
jets emanating from a central core.

Because \ls\ appeared unresolved ($\leq0.1^{\prime\prime}$) to the VLA
alone, we proceeded to study this object with milliarc sec resolution
using the VLBA at the frequency of 5 GHz (6 cm wavelength) on 8 May 1999.
The VLA in its phased array mode, equivalent to a dish of 115 m diameter,
also participated as an independent station, providing sensitive baselines
with the VLBA antennas. The source 3C345 was used as a fringe-finder,
whereas J1733$-$1304 was the phasing source for the VLA. The data were
calibrated using standard procedures in unconnected radio interferometry.
The resulting pattern of the observed visibility amplitudes, decaying as a
function of baseline length, indicated that \ls\ had structure at milliarc
sec scales.

The final synthesis map (Fig.~1) shows that bipolar jets emerge from a
central core. A deconvolved angular size of about 2 milliarc sec is
estimated for the core. The jets extend over 6 milliarc sec on the sky
oriented along a position angle (PA) of $125^{\circ}$ with respect to the
North, and they account for 20\% of the total 16~mJy flux density. To
obtain some order of magnitude estimates, we will assume that the overall
size of the radio source is approximately $6\times2$ milliarc sec$^2$.
This implies a high brightness temperature of $\sim9.4\times10^7$~K,
indicating synchrotron radiation. The \ls\ radio spectrum as a function of
frequency $\nu$, namely $S_{\nu}\propto\nu^{\alpha}$, often displays a
negative spectral index $\alpha=-0.5$ in agreement with a non-thermal
optically thin emission mechanism ({\it 3,4}). The detection of jets
occurred at a time when the source was at its typical persistent level of
radio emission, and only moderately variable, as inferred from concurrent
radio monitoring by the Green Bank Interferometer (GBI) (Fig.~2). The
absence of any precursor outburst for the radio jets strongly suggests
that they are always present and continuously emanating from the core. The
flux density ratio between the SE and NW jet components is estimated as
$2.1\pm0.4$. It seems reasonable that this brightness asymmetry reflects a
relativistic Doppler boosting effect ({\it 5}). If a continuous jet flow
is assumed, the projected velocity required is then
$v \cos{\theta}=(0.15\pm0.04)c$, where $c$ is the speed of light and
$\theta$ the ejection angle with the line of sight. It is straightforward
to then derive a lower and upper limit for the jet velocity
[$v\geq(0.15\pm0.04)c$] and the ejection angle
($\theta\leq81^{\circ}\pm2^{\circ}$), respectively.

X-ray binaries with collimated radio jets belong to the class of galactic
microquasars. The production of jets is almost certainly related to the
capture of matter from a normal star by a black hole or neutron star
companion. This is a highly energetic process with observable consequences
from radio to hard x-rays ({\it 6}) and possibly beyond. The recent third
EGRET catalog of high-energy ($E_{\gamma}>100$~MeV) $\gamma$-ray sources
({\it 7}) contains nearly 100 unidentified emitters at low galactic
latitudes. The position of \ls\ is well inside the 95\% confidence contour
of the EGRET source 3EG~J1824$-$1514, whose radius is about 0.5$^{\circ}$.
Moreover, \ls\ is the only x-ray emitter within 1$^{\circ}$ of
3EG~J1824$-$1514 listed in the ROSAT (Roentgen Satellite) All Sky Survey
({\it 8}). Such a good position agreement between an EGRET source and a
peculiar radio jet x-ray binary strongly implies that both objects are the
same. Thus, this microquasar system is likely associated with an EGRET
source. The $\gamma$-ray emission observed reveals a rather persistent
flux of $>100$ MeV photons for the last 10 years (Fig.~3).

Using modern photometric data ({\it 9}) and the reddening free parameter
formulation ({\it 10}), we obtained a distance estimate of 3.1~kpc. This
value is in excellent agreement with independent results based on the star
color excess ({\it 2}). On the other hand, a common intrinsic radio
luminosity has been recently suggested for persistent x-ray binaries ({\it
11}). According to this, the \ls\ average flux density of a few tens of
mJy at cm wavelengths would imply a rough distance value not higher than 2
kpc. Thus, different distance indicators show that \ls\ is nearby, and we
adopt a distance of 3 kpc. Therefore, this star appears to be one of the
closest, and optically brightest, microquasars among the persistent
members of this class. Several other non transient microquasars happen to
be beyond distances of about 8 kpc ({\it 12}), such as the prototypical
1E~1740.7$-$2942 in the heavily obscured regions of the Galactic Center
({\it 13}).

The synchrotron radio luminosity between 0.1 and 100~GHz for this distance
is $L_{\rm{rad}}\sim7.5\times10^{30}$~erg~s$^{-1}$. The average
$\gamma$-ray flux for all EGRET viewing periods in Fig.~3 is
$\Phi_{\gamma}=(35.2\pm6.5)\times10^{-8}$~photon~cm$^{-2}$~s$^{-1}$ with
photon spectral index $p=2.19\pm0.18$, where
$\Phi_{\gamma}\propto~E_{\gamma}^{-p}$. The EGRET photon index of \ls\ is
practically identical to that of 1E~1740.7$-$2942, i.e., steeper than the
$p<2$ values usually found for pulsars ({\it 14}). The corresponding
integrated luminosity amounts to
$L_{\gamma}(>100~\rm{MeV})\sim3.8\times10^{35}$~erg~s$^{-1}$, compared
to an x-ray luminosity ({\it 4}) of
$L_X(1.5-12~\rm{keV})\sim5\times10^{34}$~erg~s$^{-1}$. Additional
information on the source energetics can be obtained by assuming energy
equipartition between the relativistic electrons and the magnetic field
({\it 15}). We are forced to use the overall source parameters observed,
because not enough information is yet available for appropriate
calculations in the rest frame of the ejecta. The corresponding results
are nevertheless expected to be within an order of magnitude for a mildly
relativistic system. Under these assumptions, the observed radio
properties of \ls\ imply a total energy content in relativistic electrons
of $E_e\sim4.8\times10^{39}$~erg, with an equipartition magnetic field of
$\sim0.2$~G.

While flowing away into opposite jets, the relativistic electrons are
exposed to a huge output of ultraviolet (UV) photons from the hot optical
star. Thus, it appears likely that a significant fraction of the EGRET
emission arises as a result of inverse Compton (IC) scattering of these
photons by the same radio-emitting electrons. The energy shift in this
process is such that $E_{\gamma}\sim\gamma_e^2E_{\rm{ph}}$, where the
energies of the $\gamma$-ray and the stellar photon are related through
the squared Lorentz factor of the relativistic electron. For an O7 main
sequence star, a UV luminosity of $L_*\sim10^{38}$~erg~s$^{-1}$ is
expected to be mostly emitted by photons with $E_{\rm{ph}}\sim10$~eV. In
order to scatter them into $\gamma$-rays with $E_{\gamma}\sim100$~MeV,
electrons with Lorentz factors $\sim10^3$, equivalent to energies of
$\sim10^{-3}$~erg, are needed. Considering the persistent EGRET
luminosity, the lifetime of such electrons against dominant IC losses will
be $t_c\sim~E_e/L_{\gamma}\sim1.3\times10^4$~s.

The electron energy will decay with time by IC scattering according to
({\it 15}) (in centimeter-gram-second units):
\begin{equation}
\left({dE\over{dt}}\right)_{\rm{IC}}=3.97\times10^{-2}U_{\rm{rad}}E^2
\label{losses}
\end{equation}
where $U_{\rm{rad}}$ is the UV radiation energy density. For electrons
flowing away into jets, assumed perpendicular to the plane of a circular
orbit with radius $r$, we have $U_{\rm{rad}}=L_*/4\pi{c}(r^2+v^2t^2)$ at a
time $t$ after injection. For an electron with initial energy $E_0$, its
IC lifetime can be expressed as $t_c=25.2/U_{\rm{rad}}E_0$ when injected
into the jet basis close to the compact object. This implies then that the
$\gamma$-ray-emitting electrons must be initially exposed to
$U_{\rm{rad}}\sim2.0$~erg~cm$^{-3}$. Such values of radiation energy
density are available if the jets originate at a distance
$r\sim1.2\times10^{13}$~cm from the star.

Equation~\ref{losses} can be solved to give: 
\begin{equation}
E(t)={E_0\over1+(r/vt_c)\arctan{(vt/r)}}.
\label{sol}
\end{equation}
By imposing the condition that electrons with $E_0\sim10^{-3}$~erg are
able to abandon the region of heavy IC emission in the star vicinity, the
condition $\pi{r}/2vt_c<1$ must be fulfilled so that they still retain
enough energy to power the extended radio jets. This requirement allows us
to constrain the jet velocity to values $v>0.05c$, in agreement with the
previous discussion of Doppler boosting. The true jet velocity is not
likely to exceed a mildly relativistic value $v\sim0.4c$, which we crudely
estimate assuming that the 6 milliarc sec extended jets have to be
replenished in a $t_c$ time. The Lorentz factor of the jets would then be
$\gamma_v=1/\sqrt{1-(v/c)^2}\sim1.1$, i.e., not extremely relativistic.
The size of the region where $\gamma$-rays are produced in this scenario
is $vt_c>1.8\times10^{13}$~cm, i.e., larger than the orbital radius.

The central engine in \ls\ must be supplying
$\dot{E}_e\sim{L_{\gamma}}\sim3.8\times10^{35}$~erg~s$^{-1}$ in the form
of relativistic electrons. Their energy distribution is expected to be a
power law $kE^{2\alpha-1}dE=kE^{-2}dE$ to produce the observed spectral
index. Assuming electron energies in the range
$m_ec^2\leq~E~\leq\gamma_{\rm{max}}m_ec^2$ we have
$\dot{E}_e=k\int{E^{2\alpha-1}EdE}=k\ln{\gamma_{\rm max}}$. Therefore, if
the proton mass is $m_p\simeq 1800 m_e$ for every relativistic electron,
the proton mass flow into the jets can be written as
$\dot{M}_{\rm{jet}}=m_pk\int{E^{2\alpha-1}dE}\simeq1800\dot{E}_e/c^2\ln{\gamma_{\rm{max}}}\sim1.3\times10^{-9}$~$M_{\odot}$~yr$^{-1}$.
The equivalent kinetic luminosity is
$L_K=(\gamma_v-1)\dot{M}_{\rm{jet}}c^2\sim10^{37}$~erg~s$^{-1}$, which is
weakly dependent on the maximum energy cutoff assumed
($\gamma_{\rm{max}}\sim10^4$). This kinetic power is about four orders of
magnitude less than that estimated for the strong ejections of the
superluminal microquasar GRS~1915+105 ({\it 16}). Both the mass outflow
and kinetic energy estimates would not be significantly affected if
positrons are considered instead of protons, because the relativistic mass
of an electron with Lorentz factor $\sim10^3$ is comparable to $m_p$.

\ls\ is one of the nearest microquasars to be discovered. It has strong
high-energy $\gamma$-ray emission, which sets limits on the likely
velocity of its jets via inverse Compton energy losses. Most of known
microquasars were discovered only after undergoing a noticeable outburst
that triggered detection by the battery of satellites and ground-based
observatories. Some recent examples include CI Camelopardalis ({\it 17})
and the nearby transient V4641 Sagittarii ({\it 18}). The microquasar
nature of these two objects is tantalizing in that both are bright optical
stars. CI Camelopardalis was even catalogued as a variable star before its
outburst. Therefore, a careful examination of modern archive databases may
reveal a previously unnoticed population of microquasars. Indeed, our
identification of \ls\ as a potential candidate resulted from a systematic
cross-correlation between public archives of astrophysical data in the
x-ray, radio and optical domains ({\it 8, 19, 20}). The success of this
approach for systematic identification opens the possibility of new
findings which may confirm that the microquasar phenomenon is not as rare
as it seems.

\newpage

\noindent
{\bf References and Notes}

\parindent=0truecm \hangindent=0.6truecm \hangafter=1
1. C. B. Stephenson and N. Sanduleak,
{\it Publ. Warner \& Swasey Obs.} {\bf 1}, 1 (1971).

\parindent=0truecm \hangindent=0.6truecm \hangafter=1
2. C. Motch, F. Haberl, K. Dennerl, M. Pakull, E. Janot-Pacheco,
{\it Astron. Astrophys.} {\bf 323}, 853 (1997).

\parindent=0truecm \hangindent=0.6truecm \hangafter=1
3. J. Mart\'{\i}, J. M. Paredes, M. Rib\'o,
{\it Astron. Astrophys.} {\bf 338}, L71 (1998).

\parindent=0truecm \hangindent=0.6truecm \hangafter=1
4. M. Rib\'o, P. Reig, J. Mart\'{\i}, J. M. Paredes,
{\it Astron. Astrophys.} {\bf 347}, 518 (1999).

\parindent=0truecm \hangindent=0.6truecm \hangafter=1
5. T. J. Pearson and J. A. Zensus,
in {\it Superluminal Radio Sources}, J.A. Zensus and T.J. Pearson, Eds.
(Cambridge Univ. Press, Cambridge and New York, 1987), pp. 1-11.

\parindent=0truecm \hangindent=0.6truecm \hangafter=1
6. I. F. Mirabel and L. F. Rodr\'{\i}guez,
{\it Annu. Rev. Astron. Astrohys.} {\bf 37}, 409 (1999).

\parindent=0truecm \hangindent=0.6truecm \hangafter=1
7. R. C. Hartman {\it et al.},
{\it Astrophys. J. Suppl. Ser.} {\bf 123}, 79 (1999).

\parindent=0truecm \hangindent=0.6truecm \hangafter=1
8. W. Voges {\it et al.},
{\it Astron. Astrophys.} {\bf 349}, 389 (1999).

\parindent=0truecm \hangindent=0.6truecm \hangafter=1
9. J. F. Lahulla and J. Hilton,
{\it Astron. Astrophys. Suppl. Ser.} {\bf 94}, 265 (1992).

\parindent=0truecm \hangindent=0.6truecm \hangafter=1
10. B. C. Reed,
{\it Publ. Astron. Soc. Pacific} {\bf 105}, 1465 (1993).

\parindent=0truecm \hangindent=0.6truecm \hangafter=1
11. R. P. Fender and M. A. Hendry,
{\it Mon. Not. R. Astron. Soc.}, in press.

\parindent=0truecm \hangindent=0.6truecm \hangafter=1
12. J. Greiner,
in {\it Cosmic Explosions}, Proceedings of the 10th Annual Astrophysics
Conference in Maryland, College Park, MD, 11 to 13 October 1999,
S. Holt and W. W. Zhang, Eds., in press.

\parindent=0truecm \hangindent=0.6truecm \hangafter=1
13. I. F. Mirabel, L. F. Rodr\'{\i}guez, B. Cordier, J. Paul, F. Lebrun,
{\it Nature} {\bf 358}, 215 (1992).

\parindent=0truecm \hangindent=0.6truecm \hangafter=1
14. M. Merck {\it et al.},
{\it Astron. Astrophys. Suppl. Ser.} {\bf 120}, 465 (1996).

\parindent=0truecm \hangindent=0.6truecm \hangafter=1
15. A. G. Pacholczyk,
{\it Radio Astrophysics} (Freeman, San Francisco, CA, 1970).

\parindent=0truecm \hangindent=0.6truecm \hangafter=1
16. I.F. Mirabel and L. F. Rodr\'{\i}guez, 
{\it Nature} {\bf 371}, 46 (1994).

\parindent=0truecm \hangindent=0.6truecm \hangafter=1
17. R. M. Hjellming and A. M. Mioduszewski, 
{\it Int. Astron. Union Circ. 6872} (1998).

\parindent=0truecm \hangindent=0.6truecm \hangafter=1
18. R. M. Hjellming {\it et al.},
{\it Int. Astron. Union Circ. 7265} (1999).

\parindent=0truecm \hangindent=0.6truecm \hangafter=1
19. J. J. Condon {\it et al.},
{\it Astron. J.} {\bf 115}, 1693 (1998).

\parindent=0truecm \hangindent=0.6truecm \hangafter=1
20. B. M. Lasker {\it et al.},
{\it Astron. J.} {\bf 99}, 2019 (1990).

\parindent=0truecm \hangindent=0.6truecm \hangafter=1
21. D. J. Helfand, S. Zoonematkermani, R. H. Becker, R. L. White,
{\it Astrophys. J. Suppl. Ser.} {\bf 80}, 211 (1992).

\parindent=0truecm \hangindent=0.6truecm \hangafter=1
22. This research was partly supported by the Direcci\'on General de
Ense\~nanza Superior e Investigaci\'on Cient\'{\i}fica (PB97-0903) in
Spain. MR acknowledges receipt of a fellowship from Generalitat de
Catalunya. MM and JM also acknowledge support by the European Commission's
Training and Mobility of Researchers program and Junta de Andaluc\'{\i}a,
respectively. The VLBA and VLA are operated by the National Radio
Astronomy Observatory (NRAO) with Associated Universities Inc. and are
funded by the NSF. The GBI is a facility of the NSF operated by NRAO with
support from the NASA High-Energy Astrophysics program.

\vspace{2cm}

\noindent
29 March 2000; accepted 15 May 2000

\newpage

\noindent
{\bf Fig.~1.} High-resolution radio map of the nearby star \ls\ obtained
with the VLBA and the VLA in phased array mode at 6 cm wavelength. The
presence of radio jets in this high-mass x-ray binary is the main evidence
supporting its microquasar nature. The contours shown correspond to 6, 8,
10, 12, 14, 16, 18, 20, 25, 30, 40, and 50 times 0.085~mJy~beam$^{-1}$,
the rms noise. The ellipse at the bottom right corner represents the
half-power beam width of the synthesized beam, $3.4\times1.2$ (milliarc
sec$^2$) with a PA of $0^{\circ}$. The map is centered at the \ls\
position $\alpha_{\rm{J2000}}=18^h26^m15.056^s$ and
$\delta_{\rm{J2000}}=-14^{\circ}50^{\prime}54.24^{\prime\prime}$. North is
at the top and East is at the left. One milliarc sec is equivalent to
$4.5\times10^{13}$~cm (3 AU) for a distance of 3~kpc.

\noindent
{\bf Fig.~2.} GBI radio monitoring of \ls, at 2.25 GHz (13 cm), during the
weeks before and after the date of our VLBA+VLA observation, indicated by
the vertical bar. No strong flaring event was recorded, suggesting that
the presence of radio jets must be a permanent feature of \ls.

\noindent
{\bf Fig.~3.} Radio and $\gamma$-ray light curves of \ls\ and
3EG~J1824$-$1514, which we propose originate in the same object. Both \ls\
and 3EG~J1824$-$1514 are consistent with a persistent level of emission
over the last decade. The fluxes plotted here are taken from the
literature and archive data ({\it 3, 4, 19, 21}). Error bars for GBI
($\pm4$ mJy) are not shown for clarity, whereas those of the VLA are
usually smaller than the symbol size.

\newpage

\begin{figure*}[htb]
\mbox{}
\vspace{15cm}
\includegraphics{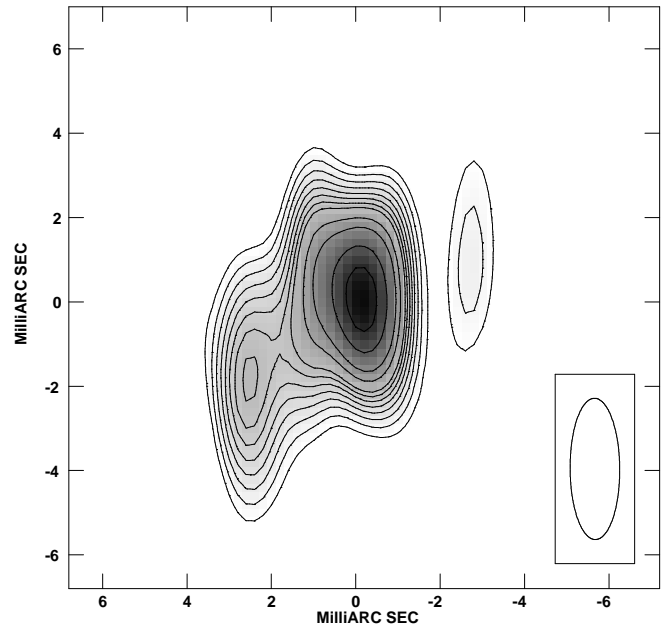}
\caption[]{J. M. Paredes {\it et al.}}
\label{f1}
\end{figure*}

\begin{figure*}[htb]
\mbox{}
\vspace{10cm}
\includegraphics{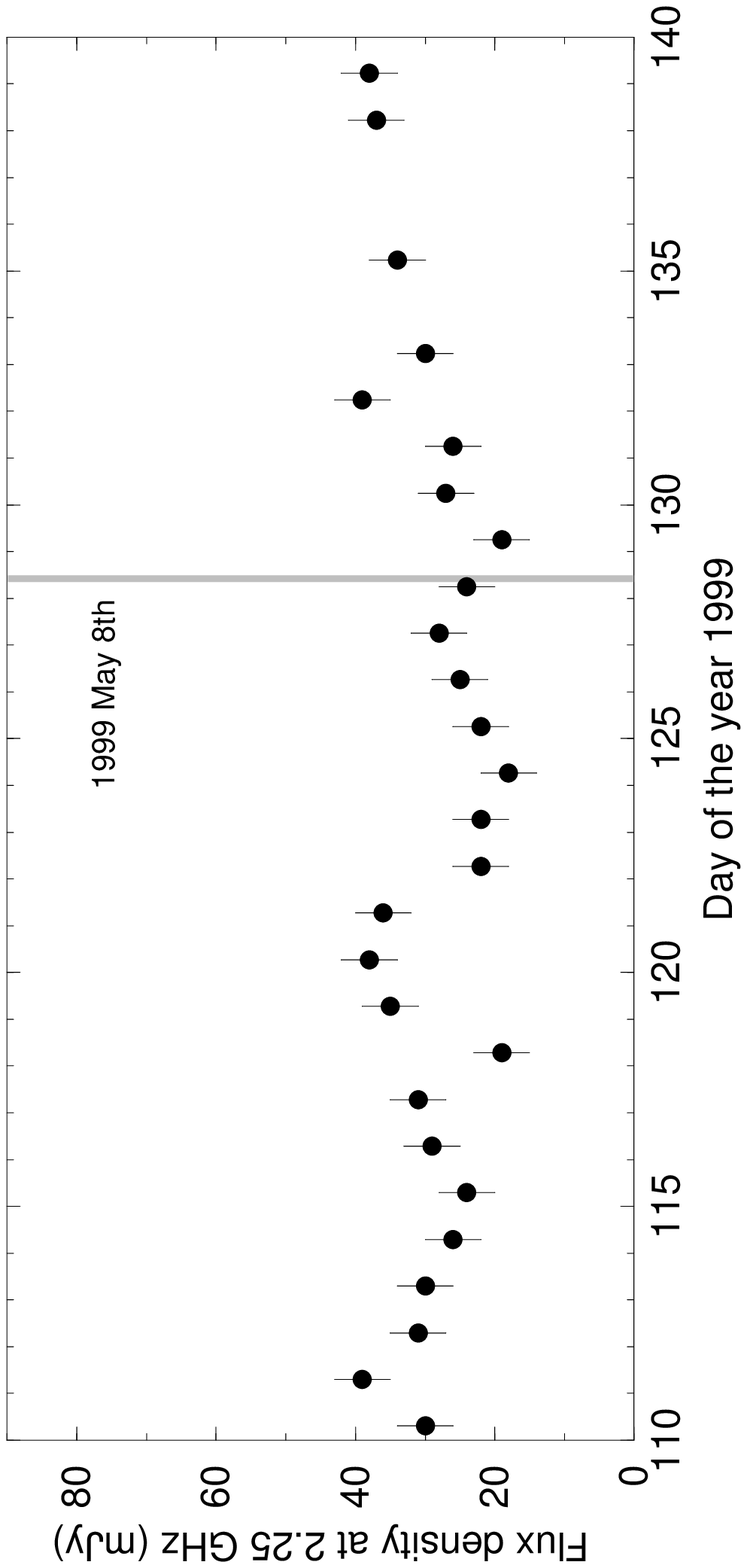}
\caption[]{J. M. Paredes {\it et al.}}
\label{f3}
\end{figure*}

\begin{figure*}[htb]
\mbox{}
\vspace{10cm}
\includegraphics{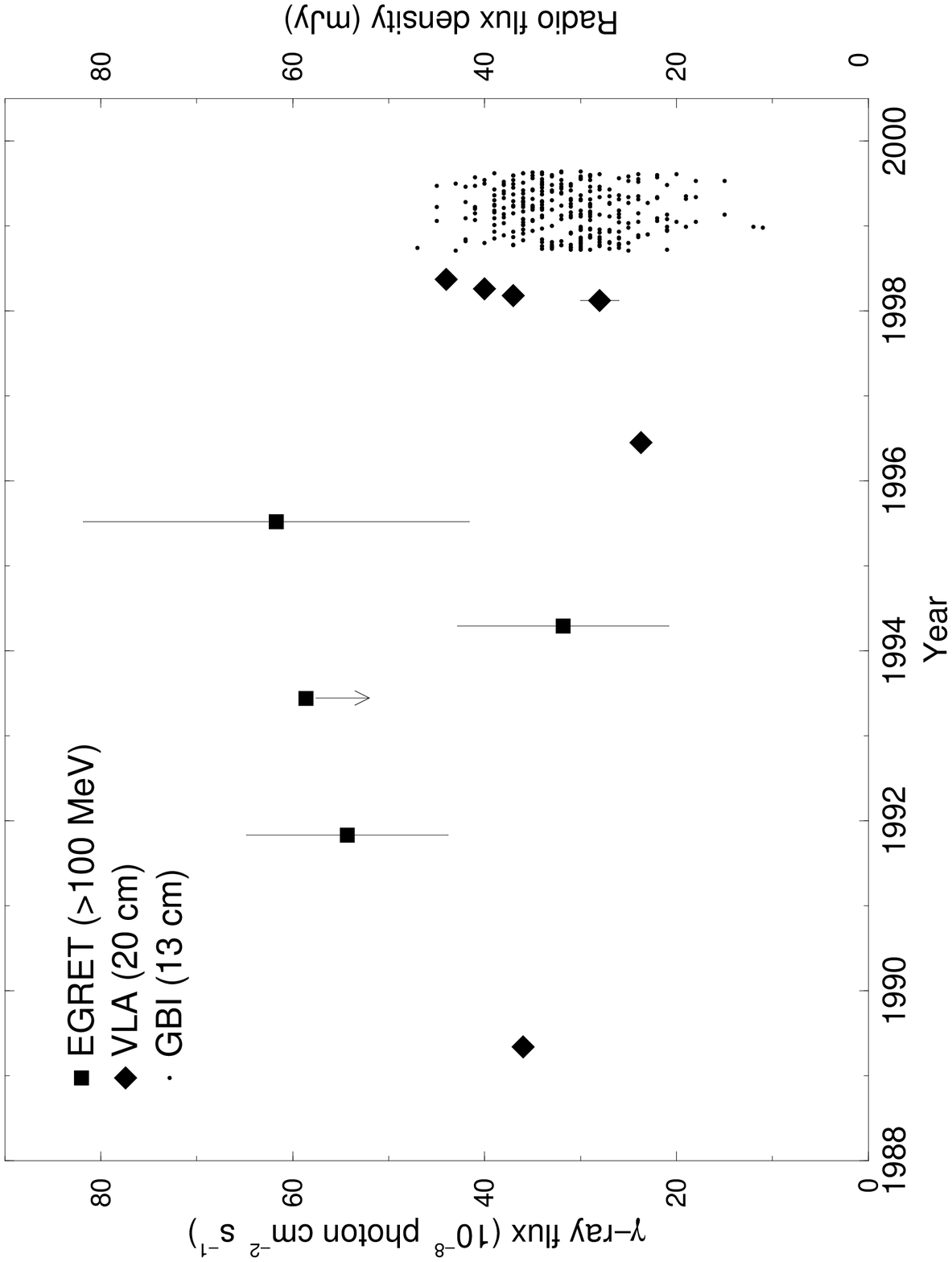}
\caption[]{J. M. Paredes {\it et al.}}
\label{f2}
\end{figure*}

\end{document}